\documentclass[aps,amssymb,article,pra,twocolumn,showpacs,showkeys,floatfix]{revtex4}

\usepackage{epsf}
\usepackage{amsmath,bm}
\usepackage{graphicx}
\begin{document}
         
  \title{Gamma and neutron radiations from condensed matter}

  \author{Boris I. Ivlev}

  \affiliation{Instituto de F\'{\i}sica, Universidad Aut\'onoma de San Luis Potos\'{\i},\\ 
  San Luis Potos\'{\i}, 78000 Mexico}

  \begin{abstract}
  Different electron states in atom are proposed. The states are bound to the electrostatic field of atomic nucleus cut off on its size. The states exist solely during acceleration of the atom exceeding
  the certain large value. The binding energy of these anomalous states is in the $10\,MeV$ range. In lead atom the transition to the anomalous state is accompanied by $33.2\,MeV$ gamma radiation. This
  is not nuclear energy. Observed high energy phenomena in lab lightning, electric explosion in liquids, and mechanical stress in solids are paradoxical since they are caused by low energy perturbations. 
  However, those observations are compatible with the electron transitions to the anomalous states since their creation requires just temporal atom acceleration but not its large kinetic energy.
  
  \end{abstract} \vskip 1.0cm

  %ralativistic wave equations, theory of quantized fields, charge distribution


  \maketitle

  \section{INTRODUCTION}
  \label{intr}
  There is the set of experimental results, which look non-explainable.
  
  In experiments \cite{OGI,OGI1} the high voltage discharge in air was revealed to produce the gamma and neutron radiations in the $10\,MeV$ range. This radiation penetrated through the
  $10\,cm$ thick lead wall. Within one discharge event the radiation elapsed approximately $10\,ns$ and corresponded to $10^{14}$ gamma quanta per second. 
  
  Since the applied voltage was less than $1\,MV$, it could not be bremsstrahlung like in X-ray tube. May be the source of the observed high energy radiation could be nuclear reactions.  But in 
  \cite{BAB} the radiation was analyzed in details and it was reasonably concluded that ``known fundamental interactions cannot allow prescribing the observed events to neutrons''. The authors of 
  \cite{OGI,OGI1} were surprised by the paradoxical inconsistency. 
  
  In experiments \cite{OGI,OGI1} it was a small power station working during $10\,ns$ (per one discharge event) and generating $100\,W$ from ``nothing'' in the form of high energy radiation in 
  the $10\,MeV$ region.
  
  In Ref.~\cite{URU} the electric explosion of titanium foils in water resulted in changes of concentration of chemical elements. The applied voltage of $\sim 10\,keV$ could not accelerate ions
  in the condensed matter up to nuclear energies required for element transmutations. Analogous results were obtained in \cite{PRI}.
  
  The surprising observations of neutrons from solids under mechanical perturbations were reported in \cite{DER,CARD2}. The phenomena in \cite{URU,PRI,DER,CARD2} are paradoxical since they are impossible 
  without high energy processes. But how these processes could be caused by the relatively low applied voltage in \cite{URU} or by the conventional (low energy) mechanical perturbations in 
  \cite{DER,CARD2}? The common feature of all above experiments was a strong acceleration (deceleration) of atoms under the mechanical conditions. 
  
  In this paper the mechanism is proposed linking low energy macroscopic perturbation of condensed matter and generating high energy. This is not nuclear energy and fusion is out of the game. The essential 
  element of this anomalous mechanism is an extreme acceleration of atoms but not their kinetic  energy.
  
  The starting point is the electrostatic nucleus field on short distance $U(r)\simeq U(0)+U''(0)r^2/2$. It is finite due to cutting off on the nuclear radius. When the electron energy $\varepsilon$ 
  compensates $U(0)\pm m\sim -10\,MeV$, in the Dirac equation one spinor tends to be singular, proportional to $1/U''(0)r^2$. It is impossible to continue this solution to $r=0$ since there is no source, 
  like $\delta(\pmb r)$, supporting the singularity. Analogously in electrostatics the singular Coulomb field should be supported by the point charge. Thus the singular solution, proportional to $1/r^2$, 
  does not exist (Sec.~\ref{aux}).

  Under a macroscopic perturbation in condensed matter an atom can move with the velocity $\dot{\pmb\xi}(t)$. The related macroscopic displacement $\pmb\xi(t)$, varying extremely slow compared to nuclear 
  times, results in the potential $U(R)$, where $\pmb R=\pmb r-\pmb\xi$. In the frame, displaced with $\pmb\xi$, the modified form $U''(0)R^2/2+i\dot{\pmb\xi}\cdot\nabla$ appears. But in the frame, where 
  the nucleus is at rest, the drag term $i\dot{\pmb\xi}\cdot\nabla$, originating from the coordinate transformation, is compensated by adapting of the electron wave function to the moving frame (when
  $\dot{\pmb\xi}(t)$ is time independent, this corresponds just to the Lorentz transformation).
  
  Besides that usual electron state, another one (anomalous) is possible. When $\ddot{\pmb\xi}(t)\neq 0$, the electron wave function can lag behind the non-inertial frame, where the nucleus is at rest. 
  In this case the drag term survives providing the singularity cut off at $R=0$. This way the anomalous state becomes temporarily physical, when the atom acceleration $\ddot{\pmb\xi}(t)$ exceeds the 
  fluctuation background. The deep energy level of the anomalous state favors it compared to usual ones.
  
  Exceeding the fluctuation background is a strong condition, which may be fulfilled in condensed matter in extreme cases (not in every day life) such as shock waves, electric discharge, mechanical 
  stress, etc.
  
  The binding energy of this anomalous state, existing long compared to nuclear times, is in the $10\,MeV$ range. It is additional in the Dirac sea. The electron transition to the anomalous state, from 
  usual atomic one, is accompanied by the gamma radiation, which is of $33.2\,MeV$ for lead atom. This is not nuclear energy. The phenomenon corresponds to the different aspect of high energy processes 
  (Sec.~\ref{mov}). 
  
  The electron transition, releasing $\sim 10\,MeV$, can also activate the nucleus deformation modes like in fission. In this process the total energy balance allows the emission of neutrons of the $MeV$
  scale. The phenomenon resembles the neutron emission caused by high energy electrons colliding the nucleus \cite{WEE} (Sec.~\ref{nucl}). 
  
  The high energy processes in condensed matter, occurring under low energy macroscopic perturbations, cannot be explained by a combination of known effects including nuclear reactions. It is argued here 
  that such processes in experiments \cite{OGI,OGI1,URU,PRI,DER,CARD2} are compatible with the electron transitions to the anomalous states. 
  
  A generic phenomenon of the $10\,MeV$ quanta radiation is expected to occur, when the ions of a relatively low energy of $100\,eV$ in a beam or a high-current glow discharge collide a properly adjusted 
  target. This is the soft scenario instead of the struggle for fusion ignition \cite{ABU} (Sec.~\ref{beam}). 
  
  In the phenomenon of sonoluminescence the surface of the collapsing bubble collides atoms of the gas inside it \cite{PUT,BRE,YOU}. The atoms acquire the velocity $\dot\xi\sim 10^3m/s$ providing 
  conditions for anomalous states at the nuclei of the gas atoms. The expected electromagnetic radiation constitutes a different (anomalous) mechanism of sonoluminescence, which is not underlain by a 
  mechanical energy transfer from the moving bubble surface to the gas inside. In the anomalous mechanism, heating of the gas in the bubble is expected to be accompanied by high-energy (in the $10\,MeV$ 
  range) electromagnetic radiation (Sec.~\ref{sonol}). 
  
  \section{AUXILIARY SOLUTIONS}
  \label{wave}
  The central potential well $U(r)$ is supposed to satisfy the condition of harmonic oscillator $U(r)\simeq U(0)+U''(0)r^2/2$ at $r\rightarrow 0$. An atomic electron is acted by the nucleus electrostatic 
  field produced by the electric charge $Ze$. The nuclear charge density is supposed to be spherically symmetric and homogeneously distributed within the sphere of the radius $r_N$ \cite{BAR}. In this case 
  \begin{equation}
  \label{25}
  U(r)=
  \begin{cases}
  -Ze^2/r,& r_N<r\\
  -3Ze^2/2r_N+\lambda r^2, & r<r_N,
  \end{cases}
  \end{equation}
  where $\lambda=Ze^2/2r^{3}_{N}$. The radiative correction to the Coulomb field (due to vacuum polarization) $(2e^2/3\pi\hbar c)\ln(0.24\hbar/mcr)$ \cite{LANDAU2} is negligible at $r\sim r_N$. As 
  shown below, short distances are mainly significant, whereas an influence of other atomic electrons is minor. 

  For deuteron ($Z=1$) the nuclear radius is $r_N\simeq 2.14\times 10^{-15}m$ and $U(0)=-3Ze^2/2r_N\simeq -1.009\,MeV$. For oxygen $^{16}{\rm O}$ ($Z=8$) the nuclear radius is   
  $r_N\simeq 2.7\times 10^{-15}m$ and $U(0)\simeq -6.4\,MeV$. For iron $^{56}{\rm Fe}$ ($Z=26$) the nuclear radius is $r_N\simeq 3.73\times 10^{-15}m$ and $U(0)\simeq -15.0\,M  eV$. For xenon 
  $^{131}{\rm Xe}$ ($Z=54$) the nuclear radius is $r_N\simeq 4.78\times 10^{-15}m$ and $U(0)\simeq -24.4\,MeV$. For lead $^{207}{\rm Pb}$ ($Z=82$) the nuclear radius is 
  $r_N\simeq 5.49\times 10^{-15}m$ and $U(0)\simeq -32.2\,MeV$. For thorium $^{228}{\rm Th}$ ($Z=90$) the nuclear radius is $r_N\simeq 5.75\times 10^{-15}m$ and $U(0)\simeq -33\,MeV$. This nucleus, 
  with the half-life of 1.92 years, emits $\alpha$-particle.
  
  \subsection{Auxiliary solutions of the Dirac equation}
  \label{aux}
  The Dirac equation has the form \cite{LANDAU2}
  \begin{equation}
  \label{213g} 
  \left[\gamma^{\mu}\left(i\hbar\partial_{\mu}-\frac{e}{c}A_{\mu}\right)-mc\right]\psi=0,
  \end{equation}
  where the bispinor $\psi$ and $\gamma$-matrices are
  \begin{equation}
  \label{212h} 
  \psi=
  \begin{pmatrix}
  \Phi\\
  \Theta 
  \end{pmatrix},
  \hspace{0.5cm}\pmb\gamma=
  \begin{pmatrix}
  0&\pmb\sigma\\
  -\pmb\sigma&0
  \end{pmatrix},
  \hspace{0.5cm}\gamma^0=
  \begin{pmatrix}
  1&0\\
  0&-1
  \end{pmatrix}.
  \end{equation}
  Here $\pmb\sigma$ is the Pauli matrix, $\partial_{\mu}=\left(\partial/c\partial t,\nabla\right)$, and $eA_{\mu}=\left(U,-e\pmb A\right)$. At $\hbar=c=1$ and $\pmb A=0$
  \begin{equation}
  \label{212g} 
  \left\{\gamma^0\left[i\frac{\partial}{\partial t}-U(r)\right]+i\pmb\gamma\cdot\nabla -m\right\}\psi(t,\pmb r)=0.
  \end{equation}
  
  For the spinor eigenfunction $\Phi(t,\pmb r)=\Phi_{\varepsilon}(\pmb r\,)\exp(-i\varepsilon t)$ and analogously $\Theta(\pmb r,t)$ 
  \begin{eqnarray}
  \label{1} 
  &&\left[\varepsilon-U(r)\right]\Phi_{\varepsilon}+i\pmb\sigma\cdot\nabla\Theta_{\varepsilon}=m\Phi_{\varepsilon}\\
  \label{2}
  &&\left[\varepsilon-U(r)\right]\Theta_{\varepsilon}+i\pmb\sigma\cdot\nabla\Phi_{\varepsilon}=-m\Theta_{\varepsilon}.
  \end{eqnarray}
  One can express $\Phi_{\varepsilon}$ from (\ref{1}) and insert into Eq.~(\ref{2}). It follows that
  \begin{equation}
  \label{6}
  \Phi_{\varepsilon}(\vec r\,)=-\frac{i\pmb\sigma\cdot\nabla\Theta_{\varepsilon}(\pmb r\,)}{\varepsilon-U(r)-m}
  \end{equation}
  and the equation for the spinor $\Theta_{\varepsilon}$, if to introduce the function $q(r)=\varepsilon-U(r)-m$, is
  \begin{equation}
  \label{7}
  -\nabla^2\Theta_{\varepsilon}+\frac{\nabla q}{q}\cdot(\nabla\Theta_{\varepsilon}-i\pmb\sigma\times\nabla\Theta_{\varepsilon})+m^2\Theta_{\varepsilon} =(\varepsilon-U)^2\Theta_{\varepsilon}.
  \end{equation}

  The spinor $\Theta_{\varepsilon}$ is chosen isotropic. This choice is possible since $\nabla q(\pmb\sigma\times\nabla\Theta_{\varepsilon})$ is proportional to the orbital momentum 
  $\pmb r\times(-i\nabla)$ that is zero $c$-number for isotropic state. See also Sec.~\ref{ful}. Since $U(r)$ is also isotropic, there is no term $\pmb\sigma\times\nabla\Theta_{\varepsilon}$ in 
  (\ref{7}) and this equation takes the form
  \begin{equation}
  \label{8}
  -\frac{q}{r^2}\frac{\partial}{\partial r}\left(\frac{r^2}{q}\frac{\partial\Theta_{\varepsilon}}{\partial r}\right)+m^2\Theta_{\varepsilon} =(\varepsilon-U)^2\Theta_{\varepsilon}.
  \end{equation}

  Let us consider the particular case of the energy $\varepsilon=\varepsilon_b$ deep in the Dirac sea, where $\varepsilon_b=U(0)+m$. As shown in Sec.~\ref{mov}, the states in the vicinity of 
  $\varepsilon_b$ can play a significant role. At small $r$ the function $q(r)\simeq -\lambda r^2$.

  Two Dirac spinors acquire the form
  \begin{eqnarray}
  \label{10l}
  &&\Phi(t,\pmb r)=-\frac{i(\pmb\sigma\cdot\pmb r)\Theta'_{\varepsilon_b}(r)}{r[U(0)-U(r)]}\exp(-it\varepsilon_b),\\
  \label{10b}
  &&\Theta(t,\pmb r)=\Theta_{\varepsilon_b}(r)\exp(-it\varepsilon_b).
  \end{eqnarray}
  The differential equation 
  \begin{equation}
  \label{10}
  -\frac{\partial}{\partial r}\left[\frac{r^2}{U(0)-U(r)}\frac{\partial\Theta_{\varepsilon_b}}{\partial r}\right]=r^2\left[2m+U(0)-U(r)\right]\Theta_{\varepsilon_b}
  \end{equation}
  follows from (\ref{8}). One can show after a little algebra that on the short distance the total solution of (\ref{10}) consists of two independent spinor parts expanded in even and odd powers 
  of $r$
  \begin{equation}
  \label{10a}
  \Theta_{\varepsilon_b}(r)=\left(1+\frac{m\lambda}{6}r^4+...\right)c_0+r\left(1+\frac{m\lambda}{10}r^4+...\right)c_b.
  \end{equation}
  Here $c_0$ and $c_b$ are constant spinors. On the large distance there are free particle solutions $\sin(r\sqrt{\varepsilon^{2}_{b}-m^2})/r$ and $\cos(r\sqrt{\varepsilon^{2}_{b}-m^2})/r$. Here 
  the Coulomb phases \cite{LANDAU2}, proportional in physical units to 
  \begin{equation}
  \label{10aa}
  \int^{r}_{0}\frac{dr_1}{\hbar c}\,U(r_1),
  \end{equation}
  are omitted.
  
  The first part in (\ref{10a}) corresponds to the usual electron state in the Dirac sea with $\Phi_{\varepsilon_b}\sim i\pmb\sigma\cdot\pmb r$. The second part in (\ref{10a}), as follows from 
  (\ref{10l}) and (\ref{10}), is the short distance limit of 
  \begin{equation}
  \label{99} 
  \Phi_{\varepsilon_b}=\frac{i\pmb\sigma\cdot\pmb r}{r^3}\frac{r^{2}_{N}}{U(0)}c_b
  \begin{cases}
  -3,& 0<r\ll r_N\\
  \beta_1rp_b\sin (rp_b+\beta_2),& r_N\ll r
  \end{cases}
  \end{equation}
  \begin{equation}
  \label{99a} 
  \Theta_{\varepsilon_b}=c_b
  \begin{cases}
  r,& 0<r\ll r_N\\ 
 (\beta_1r^{2}_{N}/r)\cos (rp_b+\beta_2),& r_N\ll r
  \end{cases}
  \end{equation}
  where $p_b=\sqrt{\varepsilon^{2}_{b}-m^2}$. In the physical units the length scale $1/p_b\sim r_N\hbar c/Ze^2$. We consider $m\ll U(0)$. The parameters $\beta_{1,2}$ are determined by the exact 
  solution of (\ref{10}) matching two asymptotics. Strictly speaking, the crossover of two asymptotics in (\ref{99}) and (\ref{99a}) occurs more complicated but we do not consider here these details.
  
  Similarly at $\varepsilon=\varepsilon_a=U(0)-m$
  \begin{equation}
  \label{99b} 
  \Phi_{\varepsilon_a}=c_a
  \begin{cases}
  r,& r\ll r_N\\
  (\alpha_1r^{2}_{N}/r)\cos (rp_a+\alpha_2),& r_N\ll r
  \end{cases}
  \end{equation}
  \begin{equation}
  \label{99c} 
  \Theta_{\varepsilon_a}=\frac{i\pmb\sigma\cdot\pmb r}{r^3}\frac{r^{2}_{N}}{U(0)}c_a
  \begin{cases}
  -3,& r\ll r_N\\
  \alpha_1rp_a\sin (rp_a+\alpha_2),& r_N\ll r
  \end{cases}
  \end{equation}
  
  At small $r$, according to (\ref{99}), $\Phi_{\varepsilon_b}\sim i\pmb\sigma\cdot\nabla(1/r)$. This term, being inserted into (\ref{2}), produces $\nabla^2(1/r)=-4\pi\delta(\pmb r)$ in that 
  equation. The $\delta$-term, supporting the singularity, does not exist and hence $c_b=c_a=0$. Analogously, in electrodynamics the singular Coulomb potential should be supported by the point charge.
  
  However (\ref{99}) - (\ref{99c}) at $r\neq 0$ can be considered as auxiliary solutions for the certain physical states if some supporting term appears in the Dirac equations resulting in finite
  $c_b$ and $c_a$. As shown in Sec.~\ref{mov}, such term exists under acceleration of the nucleus. 

  \subsection{Different types of nuclear potential}
  \label{did}
  The condition of isotropic potential $U(r)$ is not a crucial aspect. When $U(\pmb r)-U(0)\sim \alpha x^2+\beta y^2+z^2$ close to the minimum of $U(\pmb r)$, the spinor 
  \begin{equation}
  \label{10c}
  \Theta_{\varepsilon_b}=r\left[a(\theta,\varphi)+i\pmb b(\theta,\varphi)\cdot\pmb\sigma\right]+...
  \end{equation}
  is also expanded in odd powers of $r$ as in (\ref{10a}). Forms of the spinor functions $a(\theta,\varphi)$ and $\pmb b(\theta,\varphi)$ follow from (\ref{7}). As in the isotropic case, the spinor 
  $\Theta_{\varepsilon_b}$ is smooth but $\Phi_{\varepsilon_b}\sim 1/[U(0)-U(\pmb r\,)]$ is also proportional to $1/r^2$. The energy $\varepsilon_b$ has the same form as above. In the isotropic case 
  ($\alpha=\beta=1$) $a=1$ and $\pmb b=0$ as in Eq.~(\ref{10a}).

  For a model of the Dirac harmonic oscillator $U(r)=m\Omega^2r^2/2$ \cite{SHIF,ALO,ESP,AKC} the results of Sec.~\ref{aux} are also valid. In this case $\varepsilon_{b,a}=\pm m$. 

  When the nucleus is proton, the nuclear charge density is linear at small $r$ \cite{ZAC} and hence the nuclear electrostatic potential satisfies the condition $\left[U(r)-U(0)\right]\sim r^3$ 
  at small $r$. Eqs.~(\ref{10l}) - (\ref{10}) are valid for this situation. Analogously to (\ref{10a}), at $0<r\ll r_N$ two solutions are
  \begin{equation}
  \label{10p}
  \Theta_{\varepsilon_b}(r)=\left[1+\frac{mU'''(0)}{45}r^5+...\right]c_4+r^2\left(1+...\right)c_5.
  \end{equation}
  The term with $c_5$ leads to $\Phi\sim 1/r^2$ (\ref{10l}) as before. At $r_N\ll r$ the solution is (\ref{99a}) but with a different phase. The solution, similar to (\ref{10p}), is expected for 
  neutron, where in the core region the charge density is similar to proton \cite{ZAC}. 

  One can conclude that at $r\neq 0$ the auxiliary solution, proportional to $1/r^2$, of the Dirac equation exists in a nucleus with a real distribution of charge density.

  \subsection{Full set of auxiliary solutions}
  \label{ful} 
   For the central potential $U(r)$ one can re-express the auxiliary solutions of (\ref{1}) and (\ref{2}) in terms of spherical spinors \cite{LANDAU2}. In this method 
  \begin{equation}
  \label{401}
  \Phi_{j,l,m}=f(r)\Omega_{jlm},\hspace{0.5cm}\Theta_{j,l,m}=(-1)^{(1+l-l')/2}g(r)\Omega_{jl'm},
  \end{equation}
  where $l=j\pm 1/2$ and $l'=2j-l$. The spherical spinors are expressed through spherical harmonics $Y_{lm}(\theta,\varphi)$ \cite{LANDAU1}
  \begin{equation}
  \label{402} 
  \Omega_{l+1/2,l,m}=\frac{1}{\sqrt{2j}}
  \begin{pmatrix}
  \sqrt{j+m}\,Y_{l,m-1/2}\\
  \sqrt{j-m}\,Y_{l,m+1/2}\\
  \end{pmatrix},
  \end{equation}
  \begin{equation}
  \label{403} 
  \Omega_{l-1/2,l,m}=\frac{1}{\sqrt{2j+2}}
  \begin{pmatrix}
  -\sqrt{j-m+1}\,Y_{l,m-1/2}\\
  \sqrt{j+m+1}\,Y_{l,m+1/2}\\
  \end{pmatrix}.
  \end{equation}
  Eqs.~(\ref{401}) - (\ref{403}) define the set of $2j+1$ states at each total angular momentum $j$.

  The functions in Eqs.~(\ref{401}) satisfy the equations \cite{LANDAU2}
  \begin{eqnarray}
  \label{404}
  &&\left[\varepsilon-U(r)-m\right]f+g'+\frac{1-\kappa}{r}g=0\\ 
  &&\left[\varepsilon-U(r)+m\right]g-f'-\frac{1+\kappa}{r}f=0,
  \label{405}
  \end{eqnarray}
  where
  \begin{equation}
  \label{406} 
  \kappa=
  \begin{cases}
  -(l+1),& j=l+1/2\\
  l,& j=l-1/2
  \end{cases}
  \end{equation}
  For the case $b$ (with the energy $\varepsilon_{b}$) in Sec.~\ref{aux} Eqs.~(\ref{404}) and (\ref{405}) take the forms
  \begin{eqnarray}
  \label{407}
  &&q(r)f+g'+\frac{1-\kappa}{r}g=0,\\ 
  &&\left[2m+q(r)\right]g-f'-\frac{1+\kappa}{r}f=0.
  \label{408}
  \end{eqnarray}
  Let us consider the case $l=j+1/2$ in (\ref{406}). The function $q(r)\simeq -\lambda r^2$ at small $r$ and Eqs.~(\ref{407}) and (\ref{408}) turn to 
  \begin{eqnarray}
  \label{408d}
  &&2m\lambda f=\frac{1}{r^2}\frac{\partial}{\partial r}\left(r^2\frac{\partial f}{\partial r}\right)-\frac{l(l+1)}{r^2}f,\\
  &&2m\lambda r^2g=\frac{\partial^2g}{\partial r^2}-\frac{(l-1)(l-2)}{r^2}g.
  \label{408a}
  \end{eqnarray}
  The solutions at small $r$ are
  \begin{equation}
  \label{409a}
  f^{(0)}(r)=\frac{m}{j+1}r^{1/2-j},\hspace{0.5cm}g^{(0)}(r)=r^{j-1/2}.
  \end{equation}
  and
  \begin{equation}
  \label{409}
  f^{(s)}(r)=\frac{{1}}{r^{3/2+j}},\hspace{0.5cm}g^{(s)}(r)=\frac{\lambda}{2-2j}r^{3/2-j}.
  \end{equation}
  The state (\ref{409a}) is conventional. The forms (\ref{409}) are generic with the auxiliary solutions (\ref{99}) - (\ref{99a}). 

  The case $b$, studied in Sec.~\ref{aux}, relates to $j=1/2$ ($\kappa=l=1$). In this case (\ref{408a}) is an analogue of Eq.~(\ref{10}) at small $r$. One can directly check 
  that at small $r$ the solutions of (\ref{408d}) and (\ref{408a}) are
  \begin{eqnarray}
  \label{408b}
  &&f^{(s)}(r)=\frac{1}{r^2}\left(1+\frac{m\lambda}{2}r^4+...\right),\hspace{0.5cm}j=\frac{1}{2}\,,\\
  &&g^{(s)}(r)=\lambda r\left(1+\frac{m\lambda}{10}r^4+...\right).
  \label{408c}
 \end{eqnarray}

  It follows from Eqs.~(\ref{401}) - (\ref{403}) that, for the case $b$ the auxiliary solutions (we remind that they are not valid at $r=0$) are
  \begin{equation}
  \label{410}
  \Phi_{1/2,1,m}=\frac{i(\pmb\sigma\cdot\pmb r\,)}{\lambda r^3}\frac{\partial\Theta_{1/2,1,m}(r)}{\partial r},\hspace{0.28cm}\Theta_{1/2,1,m}(r)=rc_b(m).
  \end{equation}
  This is equivalent to (\ref{10l}) and (\ref{10b}) and the expansion (\ref{10a}). The spinor $c_b(m)$ has the form
  \begin{equation}
  \label{411}
  c_b(1/2)=-\frac{\lambda}{\sqrt{4\pi}}\begin{pmatrix}
  1 \\
  0 \\ 
  \end{pmatrix},\hspace{0.5cm}
  c_b(-1/2)=-\frac{\lambda}{\sqrt{4\pi}}\begin{pmatrix}
  0 \\
  1 \\ 
  \end{pmatrix}.
  \end{equation}

  Analogously one can consider the angular momentum $l=j-1/2$ in (\ref{406}) corresponding to the case $a$ ($\varepsilon=\varepsilon_a$), when $g^{(s)}\sim 1/r^{3/2+j}$ and 
  $f^{(s)}\sim r^3g^{(s)}$. 
  
  \section{ANOMALOUS ELECTRON STATES}
  \label{mov} 
  \subsection{Electron states of the moving nucleus}
  \label{genappr}
  Suppose the electron in an atom to be acted by the nuclear potential $U(|\pmb r-\pmb\xi(t)|)$ localized at the time variable position $\pmb\xi(t)$. We suppose $\dot\xi\ll c$. The field 
  of other atomic electrons is not significant since it is much smaller than the $MeV$ scale. 

  One can make the change of variable $\pmb r=\pmb R+\pmb\xi(t)$ resulting in
  \begin{equation}
  \label{522f}
  \frac{\partial\psi(t,\pmb r)}{\partial t}\rightarrow\left[\frac{\partial}{\partial t}-\dot{\pmb\xi}(t)\cdot\frac{\partial}{\partial\pmb R}\right]\psi(t,\pmb R).
  \end{equation}
  In the reference frame $(t,\pmb R)$ the nucleus is at rest. The Dirac equation acquires the form ($\hbar=1$)
  \begin{equation}
  \Bigg\{\gamma^0\left[i\frac{\partial}{\partial t}-i\dot{\pmb\xi}(t)\cdot\nabla-U(\pmb R)\right]+ic\pmb\gamma\cdot\nabla-mc^2\Bigg\}\psi(t,\pmb R)=0
  \label{522b} 
  \end{equation}
  where $\nabla=\partial/\partial\pmb R$. When $\ddot\xi=0$, (\ref{522f}) corresponds to the Lorentz transformation of coordinates in the limit $\dot\xi\ll c$.

  One can also make the transformation \cite{AKH}
  \begin{equation}
  \label{522g}
  \psi(t,\pmb R)=\bigg[1+\frac{{\pmb v}(t)}{2c}\cdot
  \begin{pmatrix}
  0&\pmb\sigma\\ 
  \pmb\sigma&0
  \end{pmatrix} 
  \bigg]\psi'(t,\pmb R)
  \end{equation}
  of the wave function to the new one ($\psi'$) obeying the equation  
  \begin{eqnarray}
  \nonumber
  &&\Bigg\{\gamma^0\left[i\frac{\partial}{\partial t}-i(\dot{\pmb\xi}-\pmb v)\cdot\nabla-U(\pmb R)\right]+ic\pmb\gamma\cdot\nabla\\
  &&-mc^2\Bigg\}\psi'(t,\pmb R)=0.
  \label{522c} 
  \end{eqnarray}
  When $\dot{\pmb\xi}(t)={\rm const}$, the condition $\pmb v=\dot{\pmb\xi}$ corresponds to the Lorentz invariance. 
   
  When $\ddot{\pmb\xi}(t)\neq 0$, one can transfer to the frame, where the nucleus is at rest, following the exact procedure accounting for all orders of $\dot\xi(t)/c$ \cite{LANDAU3}. The 
  exactly determined new electron wave function $\psi'$ follows the accelerating frame \cite{SIN}. This is the usual situation. 
  
  Unexpectedly, in the accelerating frame another electron state is possible. It corresponds to the lag of the electron wave function behind the accelerating frame resulting in $\pmb v=0$ in 
  (\ref{522g}) and (\ref{522c}). This lag is impossible without an acceleration exceeding the fluctuation background (Sec.~\ref{macr}). Now the electron state is described by the wave function 
  $\psi(\pmb R,t)$ determined by (\ref{522b}). This state is referred to as {\it anomalous}. The deep energy level of the anomalous state favors it compared to usual ones.
  
  Below we consider the anomalous state. The form 
  \begin{equation}
  \label{211} 
  \psi\simeq
  \begin{pmatrix}
  \Phi(\pmb R)\\
  \Theta(\pmb R)\\ 
  \end{pmatrix}\exp\left[-it\varepsilon_b-i\int^tdt'\Delta\varepsilon(t')\right].
  \end{equation}
  satisfies Eq.~(\ref{522b}) turning to
  \begin{equation}
  \label{212a} 
  \left[\Delta\varepsilon+q(R)-i\dot{\pmb\xi}\cdot\nabla\right]\Phi=-ic\pmb\sigma\cdot\nabla\Theta
  \end{equation}
  \begin{equation}
  \left[2mc^2+\Delta\varepsilon+q(R)-i\dot{\pmb\xi}\cdot\nabla\right]\Theta=-ic\pmb\sigma\cdot\nabla\Phi,
  \label{212b}
  \end{equation}
  where  $q(R)=\varepsilon_b-U(R)-mc^2$. 

  In condensed matter experiments an atom jumps to a neighbor position so that the function $\dot{\pmb\xi}(t)$ has a peak. Whereas in the electron system the typical time is of the nuclear scale, 
  $\dot{\pmb\xi}(t)$ varies slowly with the typical time of the inverse Debye frequency $1/\omega_D\sim 10^{-13}s$. Thus the dynamics is mainly adiabatic \cite{CON,MAC} and the spinors $\Phi$ and 
  $\Theta$ depend on $t$ through an instant value of $\dot{\pmb\xi}(t)$ in (\ref{212a}) and (\ref{212b}). 

  With the transformation $\pmb r=\pmb R+\pmb\xi(t)$ ($\dot{\xi}\ll c$) the left-hand side of the equation for QED electron propagator is analogous to (\ref{522b}). The equation for photon propagator 
  acquires the small part $\dot\xi/c$, which is not essential.

  \subsection{Auxiliary solution gives rise to the physical state}
  \label{eval}
  At small $R<r_N$ the function $q(R)\simeq-\lambda R^2$ (Sec.~\ref{wave}). The spatial scale $R\sim l$ and the energy variation $\Delta\varepsilon$ of the state can be estimated comparing the terms 
  in the left-hand side of (\ref{212a}), $\lambda l^2\sim\Delta\varepsilon\sim \dot\xi/l$. In the physical units at $m\ll |\varepsilon_b|$ and with the definition 
  $\Delta\varepsilon=s\Delta\varepsilon_0$
  \begin{eqnarray}
  \label{216}
  l=r_N\left(\frac{2\hbar{\dot\xi}}{Ze^2}\right)^{1/3}\simeq 10^{-16}[10^{-3}\dot\xi (m/s)]^{1/3}(m),\\
  \Delta\varepsilon_0 =\frac{|\varepsilon_b|}{3}\left(\frac{2\hbar{\dot\xi}}{Ze^2}\right)^{2/3}\simeq 10[10^{-3}\dot\xi (m/s)]^{2/3}(keV).
  \label{216a}
  \end{eqnarray}
  The parameter $s\sim 1$ takes various values providing possible energies of the anomalous state. The adiabatic time dependence of $l$ and $\Delta\varepsilon$ follows from $\dot\xi(t)$. The expressions 
  (\ref{216}) and (\ref{216a}) weakly depend on $Z$ since $r^{3}_{N}\sim Z$.
  
  The typical length scale  should be not shorter than the Compton radius $10^{-18}m$ of the Higgs boson. As follows from the Standard Model \cite{GLA,ENG,HIG}, on shorter distances the usual concept of 
  electron mass is not valid. Thus anyway it should be $\dot\xi>10^{-3}m/s$. 
  
  At not very small $R>l$ (but still less than $r_N$) the terms with $\dot{\pmb\xi}\cdot\nabla$ and $\Delta\varepsilon$ in (\ref{212a}) can be dropped. In this case the solution of (\ref{212a}) 
  and (\ref{212b}) is given by (\ref{99}) and (\ref{99a}). 
  
  \begin{figure}
  \includegraphics[width=7.0cm]{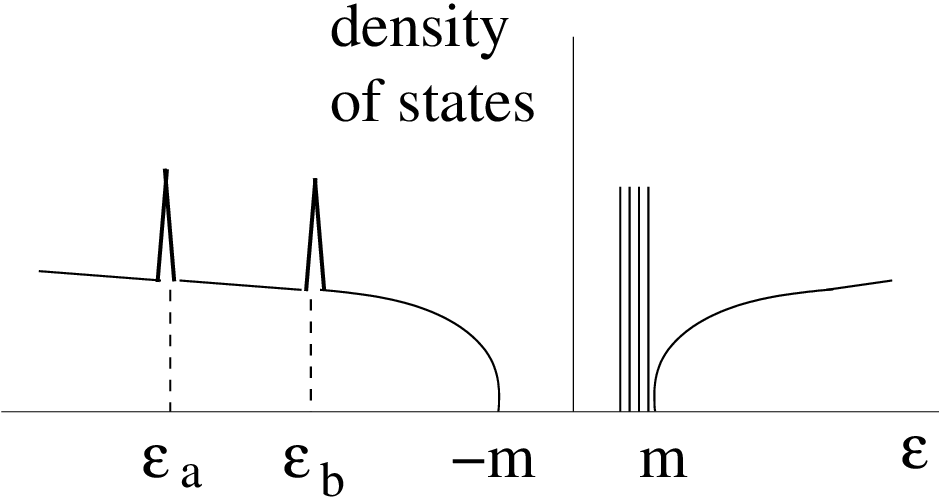}
  \caption{\label{fig1}Two peaks correspond to the anomalous state $|\varepsilon_a|\sim |\varepsilon_b|\sim 10\,MeV$. The peak width is of $10\,keV$. The peaks exist solely during nucleus acceleration.
  The usual discrete levels of the atom are shown by the vertical lines.} 
  \end{figure}

  As follows from (\ref{212a}) and (\ref{212b}), there is no singularity at $R=0$. The left-hand side of (\ref{212a}), due to finite $\dot{\pmb\xi}$, does not turn to zero 
  at $R=0$ and thus the singularity $F\sim 1/R^2$ (\ref{99}) is cut off on $R\sim l$. Details are in Appendix A. That is the auxiliary solutions of Sec.~\ref{aux} continue to $R=0$ resulting in the 
  physical states $a$ and $b$. These anomalous states, with negative energy in the $(-10)MeV$ range, are additional in the Dirac sea. The states have the width $\Delta\varepsilon\sim 10\,keV$ and are 
  shown in Fig.~\ref{fig1}.

  To study details one can approximate the anomalous wave function by
  \begin{equation}
  \label{224e} 
  \Phi\sim\frac{i\pmb\sigma\cdot\pmb R}{R^2+l^2}
  \begin{cases}
  r_N/R,& R\ll r_N\\
  \sin\left(R\sqrt{(\varepsilon_b+\Delta\varepsilon)^2-m^2}\right),& r_N\ll R
  \end{cases}
  \end{equation}
  Compared to (\ref{99}) it is put $r_Np_b\sim 1$ for simplicity. The expression analogous to (\ref{99a}) holds for $\Theta$. 
  
  The state contains energies in the interval $\Delta\varepsilon$ around $\varepsilon_b$. With the integration on $\Delta\varepsilon$ between zero and $\Delta\varepsilon\sim\Delta\varepsilon_0$
  the bispinor (\ref{211}) takes the form
  \begin{equation}
  \label{211a} 
  \psi\simeq
  \begin{pmatrix}
  F\\
  G\\ 
  \end{pmatrix}\exp\left(-it\varepsilon_b\right).
  \end{equation}
  At $r_N\ll R$ the upper spinor
  \begin{eqnarray}
  \nonumber
  &&F=\frac{i\pmb\sigma\cdot\pmb R}{R^2}\big[\exp(iR\varepsilon_b)C_b(R-ct)\\
  &&-\exp(-iR\varepsilon_b)C_b(R+ct)\big]\theta(t)
  \label{211b}
  \end{eqnarray}
  consists of the divergent and the convergent waves. The former is an outgoing wave packet but the latter tends to form a state localized at the nucleus. Such state is impossible within the
  Dirac formalism and would involve QED effects. We consider the scenario, when solely the outgoing wave is formed. This scenario is supported by the emission of unusual waves observed in experiments 
  \cite{URU,PRI1}.  
  
  The state becomes occupied at the moment $t=0$, when a usual electron falls to it during $10^{-22}s$. This is a reason of appearance of $\theta(t)$ in (\ref{211b}). The energy distribution, within
  the interval $\Delta\varepsilon$, is formed on the short distance. On the large distance it can be described by the certain distribution function. For simplicity one can use the Gaussian average
  \begin{equation}
  \label{211c} 
  C_b(R-ct)\sim\frac{1}{\sqrt{L}}\exp\left[-\frac{(R-ct)^2}{L^2}\right].
  \end{equation}
  The state is the spherical wave packet of the width
  \begin{equation}
  \label{211d} 
  L=\frac{c}{\Delta\varepsilon_0}\sim\frac{10^{-11}(m)}{[10^{-3}\dot\xi(m/s)]^{2/3}}
  \end{equation}
  and normalized for one particle. In the limit $\varepsilon_b\gg m$ the group velocity is almost $c$ and the packet smearing is weak.
  
  Summarizing, during the acceleration, when the anomalous state exists, the usual atomic electron falls to this state emitting gamma quanta with the rate of $10^91/s$ (Appendix B). After this 
  transition the electron, having the negative energy in the $(-10)MeV$ range, runs away as the propagating spherical wave (\ref{211b}). This way the atom converts into a usual ion. The arising 
  electron deficiency is compensated by electrons coming from the surrounding matter. Before and after the acceleration the atom is conventional. 
  
  \subsection{Anomalous states versus usual ones}
  \label{macr}
  In this section we present the arguments for formation of the anomalous states.
  
  Besides the macroscopic motion, describing by $\pmb\xi(t)$, the lattice site in a solid participates in the fluctuation motion $\pmb u(t)\sim 10^{-11}m$ with the typical time scale 
  $1/\omega_D\sim 10^{-13}s$ (the inverse Debye frequency) and $\langle\pmb u\rangle=0$. That is $\ddot u\sim 10^{15}m/s^2$. The fluctuating $\pmb u(t)$ is an infinite sum of Fourier harmonics
  to be averaged independently.
  
  In the limit
  \begin{equation}
  \label{224d} 
  \ddot{\pmb\xi}^{\,2}(t)\gg\langle\ddot{\pmb u}^{\,2}\rangle
  \end{equation}
  $\dot{\pmb\xi}(t)$ dominates the oscillating $\dot{\pmb u}(t)$ in the modified term $(\dot{\pmb\xi}+\dot{\pmb u})\cdot\nabla$ in Eq.~(\ref{522b}). Thus under the condition (\ref{224d}) one can 
  ignore the fluctuations considering solely the macroscopic velocity $\dot{\pmb\xi}$. In this case one can transfer to the frame, where the nucleus is at rest, following the exact procedure 
  accounting for all orders of $\dot\xi(t)/c$ \cite{LANDAU3}. The exactly determined new electron wave function $\psi'$ follows the accelerating frame \cite{SIN}. This is the usual atomic state, 
  when the nucleus moves with the acceleration $\ddot{\pmb\xi}$. 
  
  Under the condition (\ref{224d}) in the accelerating frame another electron state (anomalous) is possible. It corresponds to the lag of the electron wave function behind the accelerating frame 
  resulting in $\pmb v=0$ in (\ref{522g}) and (\ref{522c}). The anomalous electron state is described by the wave function $\psi(\pmb R,t)$ determined by (\ref{522b}). This quantum mechanical 
  state, after its formation, gets occupied with the probability (\ref{B5}) by the electron from the usual atomic state. 
  
  In the opposite limit $\ddot{\pmb\xi}^{\,2}(t)\ll\langle\ddot{\pmb u}^{\,2}\rangle$, $\dot{\pmb u}(t)$ varies fast compared to the macroscopic velocity $\dot{\pmb\xi}(t)$. In this case the natural 
  fluctuations of velocity dominate in both the change of variables (\ref{522f}) and the wave function transformation (\ref{522g}). The latter is naturally (with no lag) adapted to the arbitrarily 
  fluctuating velocity and thus the anomalous state is not formed.
  
  The criterion (\ref{224d}) is strong. To fulfill it, extreme conditions are required. In condensed matter it should be shock waves, electric discharges, dislocation motion, etc. The braking of an 
  ion beam by a properly adjusted target can be also an appropriate example. 
  
  One can show, analogously to Appendix B, that the electron-photon interaction results in a weak energy relaxation (with the rate less than $\hbar/\Delta\varepsilon$) within the energy distributed 
  anomalous state. The QED interaction does not modify the above scenario.
  
  In principle, the high energy emission can result from the usual process, namely, from a multiple absorption of low energy quanta. For example, in experiments \cite{OGI, OGI1} due to acceleration 
  the electron acquires the kinetic energy $E\simeq 0.8\,MeV$. This electron can emit the Bremsstrahlung quantum of that energy absorbed by another electron. The observed $10\,MeV$ gamma quantum 
  would correspond to the tenth order with respect to that two step process. The corresponding probability is non-physically small. 
  
  The other example of usual processes, resulted in the high energy emission, is the Meitner-Auger redistribution of the electron energy \cite{LANDAU1}. Two electrons, with the kinetic energy $E$ 
  each, due the mutual Coulomb interaction, acquire $2E$ and zero. In the high order of the perturbation theory the final electron can get a high energy. The proper probability is also non-physically 
  small. 
  
  \section{NEUTRON EMISSION}
  \label{nucl}
  In this section an excitation of nuclear collective modes by the transition to the anomalous state is studied.

  \subsection{Energy balance}
  \label{gener}
  The nucleus was treated above as a rigid object interacting via the Coulomb force with electrons. According to the liquid drop model, collective oscillations of the nuclear matter are
  possible with frequencies in a wide range on the order of $10\,MeV$ (nuclear giant resonance \cite{BOT,MIG1,BAL}). An external $\gamma$-radiation, absorbed by those collective modes, can lead to
  nuclear deformations, generic with nuclear fission, resulting in neutron emission \cite{BARB}. Nuclear collective modes correspond, for example, to ellipsoidal deformation of the spherical nucleus.

  There is another mechanism of neutron emission caused by incident high energy electrons. The direct interaction of the incident electrons with the nucleus is weaker compared to the $\gamma$-radiation. 
  However those high energy electrons can convert their kinetic energy into photons and also lead to neutron emission \cite{WEE}. 

  The perturbation theory holds with respect to the Coulomb interaction of anomalous electrons and the nuclear modes. In the electron transitions to the anomalous level these modes are directly 
  activated. In this process the electron gives up the energy $-\varepsilon_a$ to nuclear collective modes. A subsequent nucleus deformation (as in fission) can result in neutron emission analogously 
  to \cite{BARB}. 

  The absorption of the anomalous electron by the iron nucleus may, for example, correspond to the process
  \begin{equation}
  \label{224} 
  ^{56}_{26}\,{\rm Fe}+e^*\rightarrow\, ^{55}_{25}\,{\rm Mn}+n+\nu_{e}+\gamma,
  \end{equation}
  where $\nu_e$ is the electron neutrino and the symbol $e^*$ stays for the anomalous electron. We emphasize that (\ref{224}) is not an absorption of usual electron colliding the nucleus. The mass of 
  the iron nucleus is $M_{\rm Fe}\simeq 52.1028\cdot 10^3MeV$. Analogously $M_{\rm Mn}\simeq 51.1742\cdot 10^3MeV$ and $M_n\simeq 0.9395\cdot 10^3MeV$. According to these estimates, the threshold of 
  the process (\ref{224}) corresponds to the excitation (by the electron $e_A$) of the iron nucleus up to the energy of $10.45\,MeV$. In our case the excitation energy $-\varepsilon_a=15.5\,MeV$ exceeds 
  that threshold and thus the reaction (\ref{224}) is energetically possible. The emitted neutrons are expected with the energies up to $5\,MeV$. Note that the minimal excitation energy of copper or lead 
  nucleus, to emit neutrons, is around $10\,MeV$ \cite{BARB}. 

  \subsection{Electron interaction with collective nucleus modes}
  \label{mod}
  The transition rate, from the usual atomic level $A$ to the anomalous level $b$, is (\ref{B5}). In addition to this, the nucleus collective modes also interact (through the Coulomb field) with electrons. 
  In other words, ``vibrations'' of the nucleus play the analogous role as photons and thus transitions to the anomalous level can excite collective nucleus modes.

  One can start with the pure quantum mechanical description, when $A\rightarrow b$ transition occurs under the certain macroscopic perturbation $V(\pmb R,t)$. In this case the probability of the process
  is \cite{LANDAU1}
  \begin{equation}
  \label{225} 
  W= \bigg |\int dtV_{Ab}(t)\exp(-i\omega t)\bigg |^2,
  \end{equation}
  where $\omega=\varepsilon_A-\varepsilon_b$ and the matrix element is
  \begin{equation}
  \label{226} 
  V_{Ab}(t)=\int \Phi^{*}(\pmb R)V(\pmb R,t)\psi_A(R)d^3R.
  \end{equation}
  Here $\Phi(\pmb R)$ is given by (\ref{224e}) and the atomic wave function $\psi_A(R)$ is defined in Appendix B. 

  One can approximate $V(\pmb R,t)=\alpha(\pmb R)V(t)$, where the dimensionless function $\alpha(\pmb R)\sim 1$ is not zero at $R<r_N$ only and accounts for details of the Coulomb interaction with 
  nuclear deformations. In this case the probability (\ref{225}) is estimated as
  \begin{equation}
  \label{225g} 
  W\sim\frac{Zr^{4}_{N}}{a^{3}_{B}L}\int dt_1V(t_1)\exp(-i\omega t_1)\int dt_2V^{*}(t_2)\exp(i\omega t_2)
  \end{equation}

  In reality a nuclear deformation is not a macroscopic variable but a fluctuating degree of freedom. Thus one has to substitute 
  $V(t_1)V(t_2)\rightarrow\langle V(t_1)V(t_2)\rangle={\cal D}(t_1-t_2)$, where ${\cal D}(t_1-t_2)$ is the fluctuation correlator. With the Fourier component ${\cal D}_{\omega}$ of the function
  ${\cal D}(t)$ the probability (\ref{225g}) becomes linear in time $W=t/\tau_N$, where the transition rate, corresponding to neutron emission, is
  \begin{equation}
  \label{225b} 
  \frac{1}{\tau_N}\sim Z\frac{Zr^{4}_{N}}{a^{3}_{B}L}\,{\cal D}_{\omega}.
  \end{equation}
  The factor $Z$ accounts for the electron number in the atom from where the transition occurs.
  
  According to estimates of typical nuclear times and energies $V\sim\omega\sim\varepsilon_b$, the correlator ${\cal D}_{\omega}\sim\varepsilon_b$. As in Appendix B, the rate of neutron emission, 
  during the entire period of atom acceleration, is 
  \begin{equation}
  \label{225a} 
  \frac{1}{\tau_N}\sim\frac{c}{L}\frac{Z^3e^2}{\hbar c}\left(\frac{r_N}{a_B}\right)^3\sim Z^4\cdot 10^31/s.
  \end{equation}
  
  In the case of iron $1/\tau_N\sim 10^81/s$ and the electron transition to the anomalous level $a$ excites the nucleus up to the energy $\omega\simeq 16\,MeV$. This energy can trigger off the 
  process (\ref{224}) with the emission of neutrons. 

  Two substantially different phenomena, macroscopic mechanical stress in a solid and neutron emission, are hardly expected to be connected. However the concept of anomalous states links these worlds.
  
  \begin{figure}
  \includegraphics[width=7.0cm]{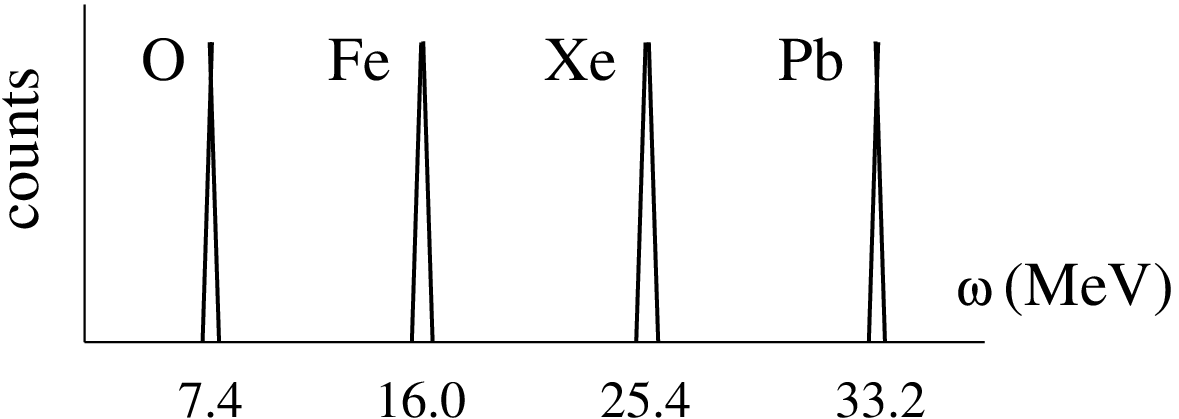}
  \caption{\label{fig2}Gamma radiation of the energy $\omega=|\varepsilon_a|+mc^2$ relates to the transitions to the anomalous level $a$ from an atomic state for different elements. The transitions to 
  the level $b$ correspond to the peaks at $\omega=|\varepsilon_b|+mc^2$ (not shown).}
  \end{figure}

  \section{EXPERIMENTS}
  \label{exp}
  Macroscopic processes in condensed matter can lead to high energy (up to tens of $MeV$) phenomena, which are of electron origin. They are connected to electron transitions to deep (anomalous) levels 
  and thus it is not nuclear energy. The necessary condition is a sufficiently strong  macroscopic acceleration (deceleration) of atoms exceeding the fluctuation background. Manifestations of these 
  phenomena are gamma radiation, with the rate (\ref{B5}), and neutron emission, with the rate (\ref{225a}). The energy spectrum of the gamma radiation is shown in Fig.~\ref{fig2}. 
  
  High energy emission under a macroscopic perturbation of condensed matter is paradoxical and cannot be explained by a combination of known mechanisms. However, there exist experimental confirmations 
  of high energy phenomena in macroscopic processes in condensed matter. 
  
  \subsection{Emission from gases}
  \label{disch}
  In high voltage discharges in gases a fast ion motion can result in conditions of anomalous states (Sec.~\ref{macr}) and thus in the high energy release. In Refs.~\cite{OGI,OGI1} the high voltage 
  discharge in air was revealed to produce a high energy radiation penetrating across the $10\,cm$ thick lead wall. It was identified as gamma and neutron radiations in the energy range of $10\,MeV$. 
  Within one discharge event the system emits $10^6$ gamma quanta during $10^{-8}s$. 
  
  Since the applied voltage was less than $1\,MV$, the bremsstrahlung energy, like in X-ray tube, could not exceed $1\,MeV$. In \cite{BAB} it was reasonably stated that known fundamental interactions 
  cannot allow prescribing the observed events to neutrons. 
  
  In experiments \cite{OGI,OGI1} it was a small power station generating $100\,W$ from ``nothing`` during $10^{-8}s$ in the form of gamma radiation in the $10\,MeV$ range 
  
  The contradictions disappear, when the anomalous electron states enter the game. According to the anomalous scenario, one atom produces $10^{9}quanta/s$ (Appendix B). This means that during the entire 
  emission process every moment of time $10^5$ ions are in the anomalous state. 
  
  \subsection{Emission from liquids}
  \label{sonol}
  Shock waves in liquids and gases are described by step like parameters in the macroscopic approach \cite{BEN}. Due to the van der Waals forces atoms of the medium start to probe the approaching shock 
  front a few Angstroms ahead of it. Since the shock velocity is about $10^3m/s$, the atoms ahead of the front acquire the same type of velocity during $10^{-13}s$. The acceleration of 
  atoms $10^{16}m/s^2$ corresponds to the condition (\ref{224d}) for creation of anomalous states. Thus the gamma radiation and neutron emission in the $10\,MeV$ range could be expected. These features 
  distinguish the anomalous phenomena and the usual acoustoluminescence \cite{KOR}.
  
  In Ref.~\cite{URU} shock waves, caused by the electric explosion of titanium foils in liquids, resulted in changes of concentration of chemical elements. Analogous results were obtained in \cite{PRI}.
  In experiments \cite{URU} the strong mechanical perturbation, leading to a large atom acceleration, was produced by the voltage of $10\,keV$. It is unlikely that the high energy processes could be 
  caused by that low voltage without the high energy processes related to transitions to anomalous states. 
  
  The neutron emission during acoustic cavitation in deuterated acetone was reported in \cite{TAL} but these results were not reproduced at other labs. See \cite{PUT1} and references therein. In 
  Ref.~\cite{LOS} X-ray radiation, caused by shock waves in water, was experimentally observed.

  The neutron emission from a deuterated medium can be supposed to be produced by nuclear processes specific for deuterons. The total mass of separate proton and neutron exceeds the mass of deuteron 
  by $2.215\,MeV$. The transition to the anomalous level releases $1.009\,MeV+m\simeq 1.52\,MeV$. This is not sufficient to break the deuteron getting free proton and neutron. Thus, if the anomalous 
  mechanism is responsible for neutron emission, the presence of deuterium is useless.
  
  In the phenomenon of sonoluminescence \cite{PUT,BRE,YOU} the surface of the collapsing bubble moves with the velocity of $0.9\times 10^3m/s$ during a few microseconds. Molecules of the gas inside 
  the bubble are collided by that supersonically moving bubble surface. Due to van der Waals forces the molecules of the gas probe the moving surface a few Angstroms ahead of it \cite{HOO}. That is the
  molecules of the gas acquire the velocity $\sim 10^3m/s$ during $(10^{-10}m)/(10^{3}m/s)\sim 10^{-13}s$ producing the acceleration of $10^{16}m/s^2$. That is the criterion (\ref{224d}) of creation of
  anomalous states, located on the gas nuclei, is expected to be fulfilled. 

  Electron transitions to the anomalous states can contribute to sonoluminescence providing the peak of gamma radiation in the $10\,MeV$ range. This high-energy radiation is an essential feature differing 
  the anomalous mechanism from the usual one with a mechanical transfer of energy to the gas from a moving bubble wall \cite{PUT,BRE,YOU}. Due to technical reasons, in \cite{PUT} the electromagnetic 
  emission could be registered in the region from $1.5\,eV$ to $6\,eV$ only. It would be amazing to detect quanta in the $10\,MeV$ range. This observation would indicate that the anomalous mechanism 
  relates to sonoluminescence.
  
  \subsection{Emission from solids}
  \label{inel}
  Macroscopic displacements of lattice sites in solids under dislocation motion or destruction under stress can result in anomalous electron states and thus in high energy processes. The examples are 
  crashing \cite{DER,CARD2}, pulling apart \cite{PUT2}, ripping and rubbing of materials.  
  
  In the experiments \cite{DER,CARD2} neutron busts were observed under strong mechanical action on solids. See also discussion and criticism of \cite{CARD2} in \cite{POM,SPA}. The experimental 
  conditions in \cite{DER,CARD2} corresponded to the motion of defects in a solid, microcracks, etc. In these processes atoms jump with the velocity $\sim 10^3m/s$ during $\sim 10^{-13}s$. This is the 
  condition of creation of anomalous states. Under those macroscopic perturbations one iron nucleus can release the total energy of $16.0\,MeV$ referred to as anomalous energy.

  There are two ways to convert that energy. Besides pure gamma radiation, one can excite nuclear degrees of freedom resulting in the fission like process of the type (\ref{224}). In this case the 
  released energy is distributed among emitted neutrons, gamma quanta, and neutrinos. The initial iron nucleus is converted into other isotope(s). The spatial concentration of such events in the solid 
  (and thus the total energy yield) depends on the probability (\ref{225a}). According to \cite{DER,CARD2}, the neutron yield under the mechanical stress essentially exceeded the natural background. 

  \subsection{Emission from ion beams}
  \label{beam}
  A moving ion in a beam (see for example \cite{SHA}) or in a high-current glow discharge accelerates a few Angstroms close to the metallic target surface due to the electrostatic mirror forces. The 
  acceleration is the electrostatic force divided by the ion mass and thus the ions should not be heavy. The ion energy can be low. These conditions are opposite to ones for fusion ignition by high 
  energy heavy ion beams \cite{ARN}. 
  
  The ion of the low kinetic energy $E\sim 100\,eV$, having thus the velocity of $\sim 10^4m/s$, accelerates during $10^{-14}s$. Resulting bremsstrahlung quanta are in the energy range of $0.1\,eV$. 
  According to Sec.~\ref{macr}, because of the acceleration of the nucleus, the anomalous state with the energy $\varepsilon_a=U(0)-m$ is expected to form. The creation of anomalous states in ion beam 
  requires the acceleration condition (\ref{224d}). In this case the fluctuation background, formed by target atoms, depends on a beam-target adjustment. The ions should be mainly reflected by the 
  target to avoid the fluctuation effect of its atoms. This condition contrasts to usual ion experiments, when the penetration inside the target is the main purpose.
  
  The parameter $U(0)$ for various nuclei is given in Sec.~\ref{wave}. The transition to that level, from 
  a usual atomic state (Appendix B), leads to the emission of the gamma quantum with the energy $\omega=-\varepsilon_a+m$. In Fig.~\ref{fig2} the corresponding emission peaks are shown for four various 
  ions. The assisting peaks for transitions to the level $b$ (not shown) are $2m\simeq 1.02\,MeV$ lower in energy.
  
  The width of each peak in Fig.~\ref{fig2} is approximately $\Delta\varepsilon$ that is $0.1\%$ of its position. This quantum mechanical width is slightly influenced by electron-photon interaction 
  (Sec.~\ref{macr}).
   
  Therefore the collision of a target by the low-energy ($E\sim 100\,eV$) ion, in a beam or a high-current glow discharge, is expected to result in the high ($\omega\sim 10\,MeV$) energy radiation. 
  The energy $\omega $ is $E$-independent. The high-energy radiation is a consequence of ion acceleration leading to the formation of anomalous states and the subsequent transitions to them. A neutron 
  emission is also possible. This follows from comparison of the rates (\ref{B5}) and (\ref{225a}).  
   
  \subsection{Energy production}
  \label{release}
  The $100\,W$ power station, producing $10\,MeV$ radiation from ``nothing'' during $10\,ns$ in experiments \cite{OGI,OGI1}, points to the possibility of energy production. 
  
  A steady ion beam, colliding the properly adjusted target, is expected to permanently reproduce the high energy emission. This is because instead of one acceleration shot there are many small shots 
  related to the target collision by individual ions in the beam. This high energy process can be referred to as radiation of anomalous energy. This is not nuclear energy despite it is in the $10\,MeV$ 
  range. 
  
  At present one can rather point on the emission spectrum in Fig.~\ref{fig2} but not on an amount of the emitted energy.
  
  \section{DISCUSSIONS}
  \label{disc} 
  Macroscopic low energy perturbations in condensed matter are not expected to activate processes of the $10\,MeV$ energy scale. However those slow varying in time perturbations can trigger off 
  electron transitions to deep lying anomalous states. These states are formed due to the lag of the electron wave function behind the non-inertial frame, where the nucleus is at rest. The anomalous 
  states are additional in the Dirac sea. They can exist solely under sufficiently large acceleration of nuclei. According to the evaluations in Sec.~\ref{macr}, this acceleration should exceed the 
  certain fluctuation level. The nucleus kinetic energy is not relevant. 
  
  Accompanying gamma radiation, in the range of $10\,MeV$, is connected to the electron processes but not to nuclear ones. So it is not nuclear energy and fusion is out of the game. The phenomenon 
  corresponds to the different aspect of high energy processes.
  
  In the electron transitions to the anomalous states the electron can also give up its energy to nuclear collective modes. A subsequent nucleus deformation, like in fission, can lead to neutron emission. 
  
  The examples of strong nucleus acceleration are the high voltage air discharge, the strong mechanical perturbation of solids, an ion beam colliding a target, a glow discharge, etc.
  
  The collision of a properly adjusted target by the low-energy ($E\sim 100\,eV$) ions in a beam is expected to result in the high energy ($\omega\sim 10\,MeV$) electromagnetic radiation. The energy 
  $\omega$ depends on type of ions but not on $E$. This way $100\,eV$ ion produces $10\,MeV$ from ``nothing'' if to forget about anomalous processes. A neutron emission is also possible. At present it is 
  hard to estimate the energy yield. Anyway, this is the soft scenario instead of the struggle for fusion ignition \cite{ABU}. 
  
  The experiments with high voltage discharge in air (lab lightning) \cite{OGI,OGI1} revealed the high energy neutron and gamma radiations in the approximately $10\,MeV$ range. The radiation penetrated 
  through the $10\,cm$ thick lead wall. The amount of $10\,MeV$ one order of magnitude exceeded the energy directly acquired by each particle in the experiment and thus nuclear reactions looked impossible. 
  In the paper \cite{BAB} it was reasonably concluded that the known fundamental interactions could not allow prescribing the observed events to neutrons in \cite{OGI,OGI1}.
  
  In spite of $10\,MeV$ quanta were not expected to appear in \cite{OGI,OGI1}, it was a small power station producing $100\,W$ in the form of those quanta and acting $10\,ns$, which is the duration of 
  the emission within each discharge event. It is highly likely that this paradoxical radiation of the high energy quanta and neutrons (anomalous radiation) was caused by electron transitions to the 
  anomalous levels. 
  
  Is that small power station a prototype of real devices based on radiation of the anomalous energy? Suppose that in the certain process $10^{18}$ atoms (approximately $0.1\,mg$ of matter) are properly 
  accelerated and contain the anomalous states. Creation of such extreme experimental conditions is a matter of study. That power station would produce $10^{15}W$ during the acceleration process. The power 
  of that station would exceed one hundred times the power of the biggest industrial power plant. For comparison, the energy of $10^{15}J$ is released under the explosion of one megaton of trotyl. 
  
  Besides fast ion experiments, acceleration of nuclei (resulting in anomalous states and thus in a high energy emission) is possible under a strong mechanical action on solids. In this case lattice 
  sites jump to new positions. For example, it could be under dislocation motion in solids. The neutron emission under the mechanical crash of solids was reported in \cite{DER,CARD2}. 
  
  In the phenomenon of sonoluminescence the surface of the collapsing bubble collides atoms of the gas inside it. The acceleration of these atoms is expected to lead to the creation of the anomalous 
  states located now on nuclei of the gas atoms. This provides another (anomalous) mechanism of sonoluminescence, which is not underlain by a mechanical energy transfer from the moving bubble surface to 
  the gas inside. Thus the conventional heating of the gas in the bubble is expected to be accompanied by gamma radiation in the $10\,MeV$ range. 
  
  \section{CONCLUSIONS}
  \label{conc}
  Macroscopic low energy phenomena in condensed matter trigger off formation of anomalous electron states. Falling to the anomalous level electrons lead to gamma radiation in the $10\,MeV$  energy 
  range. This is anomalous electron energy. This is not nuclear energy. The phenomenon corresponds to the different aspect of high energy processes. Associated excitation of nuclear collective modes 
  results in neutron emission. Thus anomalous electron states link usual macroscopic phenomena in condensed matter and the processes of the $10\,MeV$ scale. Paradoxical experimental results on gamma 
  and neutron radiations, providing energy from ``nothing'', are in agreement with the concept of anomalous states.

  \acknowledgments

  I am grateful to J. Engelfried and A. M. Loske for stimulating discussions. This work was supported by CONACYT through grant 237439. 

  \appendix

  \section{CUT OFF SINGULARITY}
  Besides the usual set of electron states in the Dirac sea there are anomalous ones. To study them suppose the nuclear displacement has the $z$-component $\xi$ only. Eqs.~(\ref{212a}) and 
  (\ref{212b}) with the notation $\pmb R=(\pmb\rho,z)$ take the forms at $R<r_N$ ($\hbar$=$c$=1) 
  \begin{equation}
  \label{A1} 
  \left(\Delta\varepsilon-\lambda R^2-i\dot{\xi}\frac{\partial}{\partial z}\right)\Phi=-i\left(\sigma_z\frac{\partial}{\partial z}+\pmb\sigma\cdot\frac{\partial}{\partial\pmb\rho}\right)\Theta,
  \end{equation}
  \begin{equation}
  2m\Theta=-i\left(\sigma_z\frac{\partial}{\partial z}+\pmb\sigma\cdot\frac{\partial}{\partial\pmb\rho}\right)\Phi
  \label{A2}
  \end{equation}

  The solution of Eqs.~(\ref{A1}) is
  \begin{equation}
  \label{A3}
  \Phi(\pmb\rho,z)=\int^{z}_{0}\frac{dz_1}{\dot\xi}A(\rho,z,z_1)\left(\pmb\sigma\cdot\frac{\partial}{\partial\pmb\rho}+\sigma_z\frac{\partial}{\partial z_1}\right)\Theta(\pmb\rho,z_1),
  \end{equation}
  where 
  \begin{equation}
  \label{A4}
  A(\rho,z,z_1)=\exp\left[\frac{i(z^3-z^{3}_{1})}{3l^3}+\frac{i(z-z_1)}{l^3}\left(\rho^2-sl^2\right)\right].
  \end{equation}
  The parameters $l$ and $\Delta\varepsilon=s\Delta\varepsilon_0$ are defined by (\ref{216}) and (\ref{216a}). The parameter $s$ (Sec.~\ref{eval}) defines the possible energies $\Delta\varepsilon$. 
  The substitution of (\ref{A3}) into (\ref{A2}) leads to the equation for $\Theta$ 
  \begin{eqnarray}
  \nonumber
  &&\left[2m\dot\xi+i\left(\sigma_z\pmb\sigma\cdot\frac{\partial}{\partial\pmb\rho}+\frac{\partial}{\partial z}\right)\right]\Theta(\pmb\rho,z)\\
  \nonumber
  &&=\int^{z}_{0}dz_1A(\rho,z,z_1)\bigg[\left(\frac{\rho^2+z^2}{l^3}-\frac{s}{l}\right)\sigma_z-i\pmb\sigma\cdot\frac{\partial}{\partial\pmb\rho}\\
  &&+\frac{2(z-z_1)}{l^3}\pmb\sigma\cdot\pmb\rho\bigg]\left(\pmb\sigma\cdot\frac{\partial}{\partial\pmb\rho}+\sigma_z\frac{\partial}{\partial z_1}\right)\Theta(\pmb\rho,z_1). 
  \label{A5}
  \end{eqnarray}
  At $R>l$ the typical $(z-z_1)$ in the exponent (\ref{A4}) $l^3/R^2$ is small resulting in locality on $z$. Thus the term $\dot\xi\partial/\partial z$ in (\ref{A1}) can be dropped and Eq.~(\ref{7}) 
  is restored at $l<R<r_N$. 
  
  At the region $R<r_N\sim 10^{-15}m$ considered, the term $2m\dot\xi$ in (\ref{A5}) can be dropped and it is clear that $\Theta(\pmb\rho,z)$ varies on $\rho,z\sim l\sim 10^{-16}m$.
  At $R>l$, according to (\ref{A1}), $\dot\xi$ does not play role and $\Theta$ consists of two contributions as in (\ref{10a}). 
  
  Those conclusions are valid at $s\sim 1$ ($\Delta\varepsilon\sim\Delta\varepsilon_0$). At $1\ll s$ the locality on $z$ occurs, when $z\gg l/s$ since $(z_1-z)\sim l/s$ in $\Theta(\pmb\rho,z_1)$. 
  When $l/s$ is below the certain minimal length $d_m$, the local equation holds in the total allowed region (outside the radius $d_m$) resulting in the solution (\ref{99}) and (\ref{99a}), which 
  correspond to the absence of the anomalous state. Thus the minimal length $d_m$ determines the line width $\Delta\varepsilon\sim\hbar c/d_m$ in Fig.~\ref{fig2}. We suppose $d_m$ to be not very 
  small fraction of the nucleus radius $r_N$ that is $\Delta\varepsilon\sim\Delta\varepsilon_0$.
  
  Suppose that there is the singularity of $\Theta(\pmb\rho,z)$ at the point $\rho=z=0$. Then close to the singularity the form of $\Theta(\pmb\rho,z)$ is determined by the most singular part of 
  Eq.~(\ref{A5})
  \begin{eqnarray}
  \nonumber
  &&\pmb\sigma\cdot\frac{\partial}{\partial\pmb\rho} \int^{z}_{z_0}dz_1
  \left(\pmb\sigma\cdot\frac{\partial}{\partial\pmb\rho}+\sigma_z\frac{\partial}{\partial z_1}\right)\Theta(\pmb\rho,z_1)\\
  &&+\sigma_z \left(\pmb\sigma\cdot\frac{\partial}{\partial\pmb\rho}+\sigma_z\frac{\partial}{\partial z}\right)\Theta(\pmb\rho,z).
  \label{A9}
  \end{eqnarray}
  The increase at $\rho\rightarrow 0,\,\,z\rightarrow 0$ of the second part in (\ref{A9}) should be compensated by the analogous increase of the first part. But in this part, due to the 
  $z'$-integration, there is the $z$-independent part, which is singular on $\rho$ only. So the compensation is impossible and thus that singularity is not formed. 
  
  Another possibility of singularity may occur, when despite the singular $\Theta$ the form $\left(\pmb\sigma\cdot\partial/\partial\pmb\rho+\sigma_z\partial/\partial z\right)\Theta$ is zero at 
  any small but finite $\rho$ and $z$ that is proportional, for example, to $\delta(\pmb\rho)\delta(z)$. Such situation may be realized, when $\Theta=\pmb\sigma\cdot\partial(1/R)/\partial\pmb R$. 
  In this case the first part in Eq.~(\ref{A9}) becomes 
  \begin{equation}
  \label{A10}
  \exp\left(\frac{iz^3}{3l^3}-\frac{iz}{l}\right)\theta(z)\,\pmb\sigma\cdot\frac{\partial\delta(\pmb\rho)}{\partial\pmb\rho}.
  \end{equation}
  where we accounted for the exponential term dropped in (\ref{A9}). The part (\ref{A10}) is singular (on $\pmb\rho$) even at $z>l$, where the smooth solution (\ref{10a}) should hold. Thus the 
  assumption of singularity is not correct and the anomalous solution, being continued to $R=0$, remains non-singular. In the case of derivatives of $\delta(z)$ one should use the proper expansion 
  on $z'/l$ in $A$ to obtain the expression similar to (\ref{A10}). 
  
  We see that the anomalous functions, determined by (\ref{99}) and (\ref{99a}) at $R>l$, are finite after continuation to $R=0$. 
  
  \section{PROBABILITY OF PHOTON EMISSION}
  The anomalous state $b$ is described by the wave function specified in Sec.~\ref{eval}. This anomalous state can be occupied by a transition (with the gamma radiation) from the usual atomic state $A$. 
  This process resembles the pair annihilation. The corresponding transition rate is (in physical units) \cite{AKH} 
  \begin{eqnarray}
  \nonumber
  &&\frac{1}{\tau_A}=Z\frac{e^2c}{4\pi}\int\frac{d^3k}{k}\langle A|\gamma^{\mu}\exp(i\pmb k\cdot\pmb R)|b\rangle\\
  &&\langle b|\gamma^{\mu}\exp(-i\pmb k\cdot\pmb R)|A\rangle\delta(mc^2-\varepsilon_b-\hbar ck).
  \label{B4} 
  \end{eqnarray}
  The factor $Z$ accounts for the number of electrons in the atomic state $A$. One estimates 
  \begin{equation}
  \label{B2}
  \int\frac{d^3k}{k}\delta(mc^2-\varepsilon_b-\hbar ck)\sim \frac{\varepsilon_b}{(\hbar c)^2}.
  \end{equation}
  The proper wave function in a heavy atom is of the type $\psi_A(R)\sim\exp(-R^2/a^{2}_{0})/a^{3/2}_{0}$. Here $a_0\sim a_B/Z^{1/3}$ \cite{LANDAU1}, where $a_B$ is the Bohr radius. From here 
  it follows that $1/R\sim k\sim (Ze^2/\hbar c)1/r_N$. One should put $|b\rangle\sim 1/(R\sqrt{L})$. The nucleus radius is $r_N=r_0Z^{1/3}$, where $r_0\sim 10^{-15}m$.

  With the above estimates 
  \begin{equation}
  \label{B5}
  \frac{1}{\tau_A}\sim\frac{\Delta\varepsilon}{\hbar}\left(\frac{\hbar c}{e^2}\right)^2\left(\frac{r_0}{a_B}\right)^3\sim 10^{9}1/s
  \end{equation}
  is the electron transition rate to the anomalous state from the usual atomic state $A$. The gamma quantum of the energy $|\varepsilon_b|+mc^2$ is emitted. After the transition the electron 
  leaves the nucleus as the outgoing wave and the new transition occurs. So that (\ref{B5}) is the rate of gamma emission during the entire period of atom acceleration.

\end{document}